\documentclass{kluwer}
\usepackage{graphicx,float}

\begin{document}
\begin{article}
\begin{opening}

\title{ In-orbit Vignetting Calibrations of XMM-Newton Telescopes }

\author{ D. H. \surname{Lumb}  \email{dlumb@rssd.esa.int}}
\institute{Adv. Concepts \& Science Payloads Office, European Space Agency, ESTEC,2200AG Noordwijk, Netherlands}
\author{A. \surname{Finoguenov}}  
\institute{Max-Planck Institut f\"ur extraterrestrische Physik, Giessenbachstra\ss e 1, D-85748 Garching, Germany}
\author{R. \surname{Saxton} }
\institute{XMM Survey Science Centre, Dept. of Physics \& Astronomy, Leicester
University, Leicester LE1 7RH, U.K. }
 \author{B. \surname{Aschenbach}} 
\institute{ Max-Planck Institut f\"ur extraterrestrische Physik, Giessenbachstra\ss e 1, D-85748 Garching, Germany}
\author{P. \surname{Gondoin}  }
\institute{Adv. Concepts \& Science Payloads Office, European Space Agency, ESTEC, 2200AG Noordwijk, Netherlands}
\author{M. \surname{Kirsch} }
\institute{XMM Science Ops. Centre, European Space Agency, Apartado -
  P.O. Box 50727, 28080 Madrid, Spain}
 \author{I. M. \surname{Stewart}}
\institute{XMM Survey Science Centre, Dept. of Physics \& Astronomy, Leicester
University, Leicester LE1 7RH, U.K. }


\begin{abstract}We describe  measurements of the mirror vignetting in the
XMM-Newton Observatory made in-orbit, using observations of SNR G21.5-09 and SNR 3C58 with 
the EPIC
imaging cameras. The instrument features that complicate these measurements
are briefly described. We show the spatial and energy dependences of
measured vignetting, outlining assumptions made in deriving the eventual
agreement between simulation and measurement. Alternate methods to confirm these
are described, including an assessment of source elongation with
off-axis angle, the surface brightness distribution of the diffuse X-ray
background, and the consistency of Coma cluster emission at different
position angles. A synthesis of these measurements leads to a change
in the  XMM calibration data base, for the optical axis of two of the
three telescopes, by in excess of 1 arcminute. This has a small but measureable
effect on the assumed spectral responses of the cameras for on-axis targets.
\end{abstract} 

\end{opening}
\section{INTRODUCTION}
\label{sect:intro}  
XMM-Newton  \cite{Jansen} comprises 3 co-aligned telescopes, each
with effective area at 1.5keV of $\sim$1500cm$^{2}$, and Full Width
Half Maximum (FWHM) angular resolution of $\sim$5 arcseconds. 
The 3 telescopes each have
a focal plane CCD imaging spectrometer camera provided by the EPIC
consortium. Two also have a reflection grating array, which splits
off half the light, to provide simultaneous high resolution dispersive
spectra. These two telescopes are equipped with EPIC MOS cameras
\cite{mos}, which are conventional CMOS CCD-based images
enhanced for X-ray sensitivity. The third employs the EPIC PN
camera \cite{pn} which is based on a pn-junction multi-linear
readout CCD.  The EPIC cameras offer a field of view (FOV) of $\sim$30
arcminute diameter, and an energy resolution of typically 100~eV
(FWHM) in the range $\sim$0.2--10~keV. The two MOS telescopes are
equipped with a Reflection Grating instrument \cite{denHerder} that
has its own dedicated readout camera.

The in-orbit calibration of the XMM-Newton mirrors has been reported
elsewhere \cite{Aschenbach}, with special reference to the on-axis angular
resolution (Point Spread Function, PSF). A second important calibration data
set that is critical for analyzing spectroscopic information is the
energy-dependent effective area \cite{Aschenbach2002}. Both these features
are under constant review as a result of improving knowledge of the
instrumentation, and the requirements imposed by new science
investigations. In this work we concentrate on different aspects of mirror performance
that must be calibrated in the context of other scientific drivers
which include, for example, cluster radial brightness
distribution for determining gas mass, exposure maps and counts-to-flux
conversions in population studies and diffuse background normalization
measurements etc.. 
The reduction in effective area with radial distance from the field of view
centre, or vignetting, must be accurately determined to support these investigations. 

\begin{figure}
\caption{Image in the MOS-2 detector plane, of the
  mean difference in the total-band (0.2-12 keV) flux seen by MOS-2 and
  MOS-1 expressed in Sigma. Bright pixels indicate an excess of flux in
  MOS-2 and dark pixels an excess in MOS-1. } \label{seren_image} 
\begin{tabular}{c}
    \includegraphics[bb = 20 100 573 740, clip,scale=0.65]{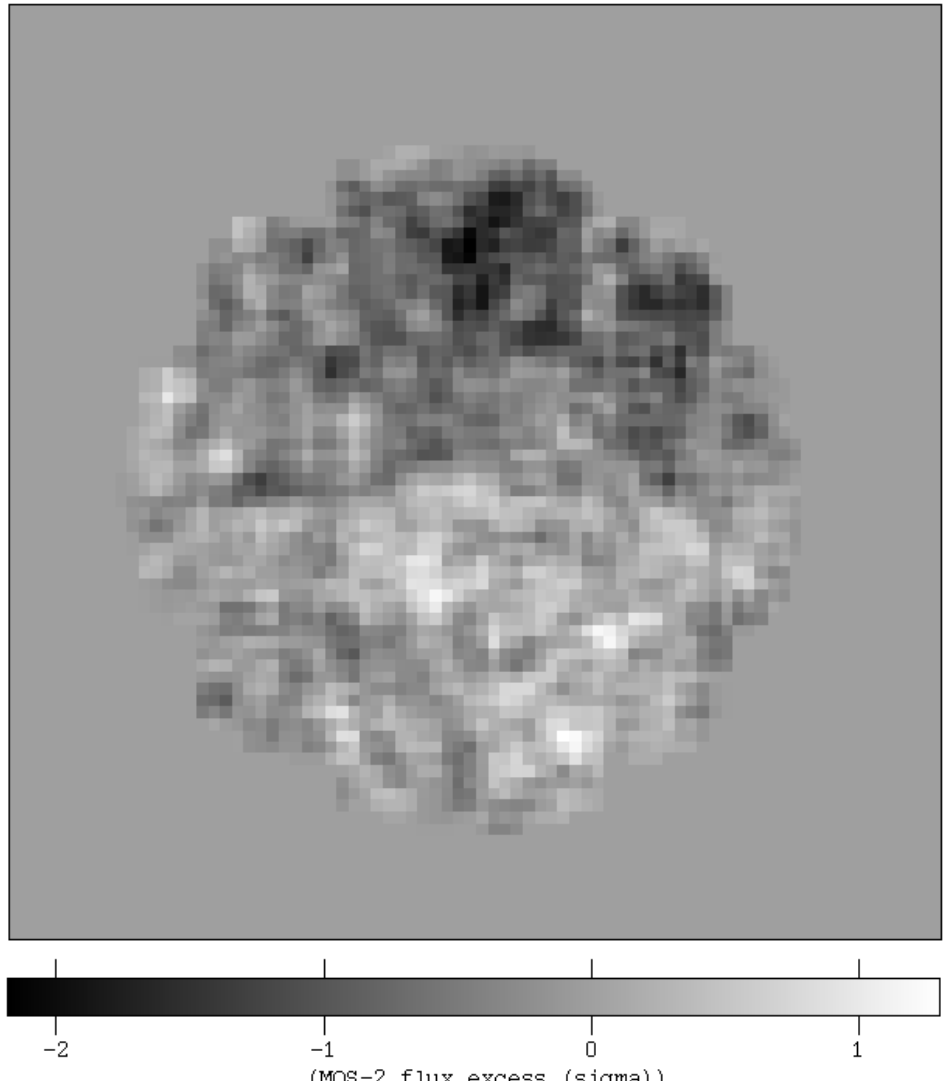}
\end{tabular}
\end{figure}

\begin{figure}
\caption{ The equivalent image for the Epic-pn detector
  plane, of the mean difference in the total-band (0.2-12 keV) flux seen by
  Epic-pn and MOS-1 expressed in Sigma. Bright pixels indicate an excess of
  flux in pn and dark pixels an excess in MOS-1. }  \label{seren_image2}
\begin{tabular}{c}
    \includegraphics[bb = 36 170 478 680, clip, scale=0.65]{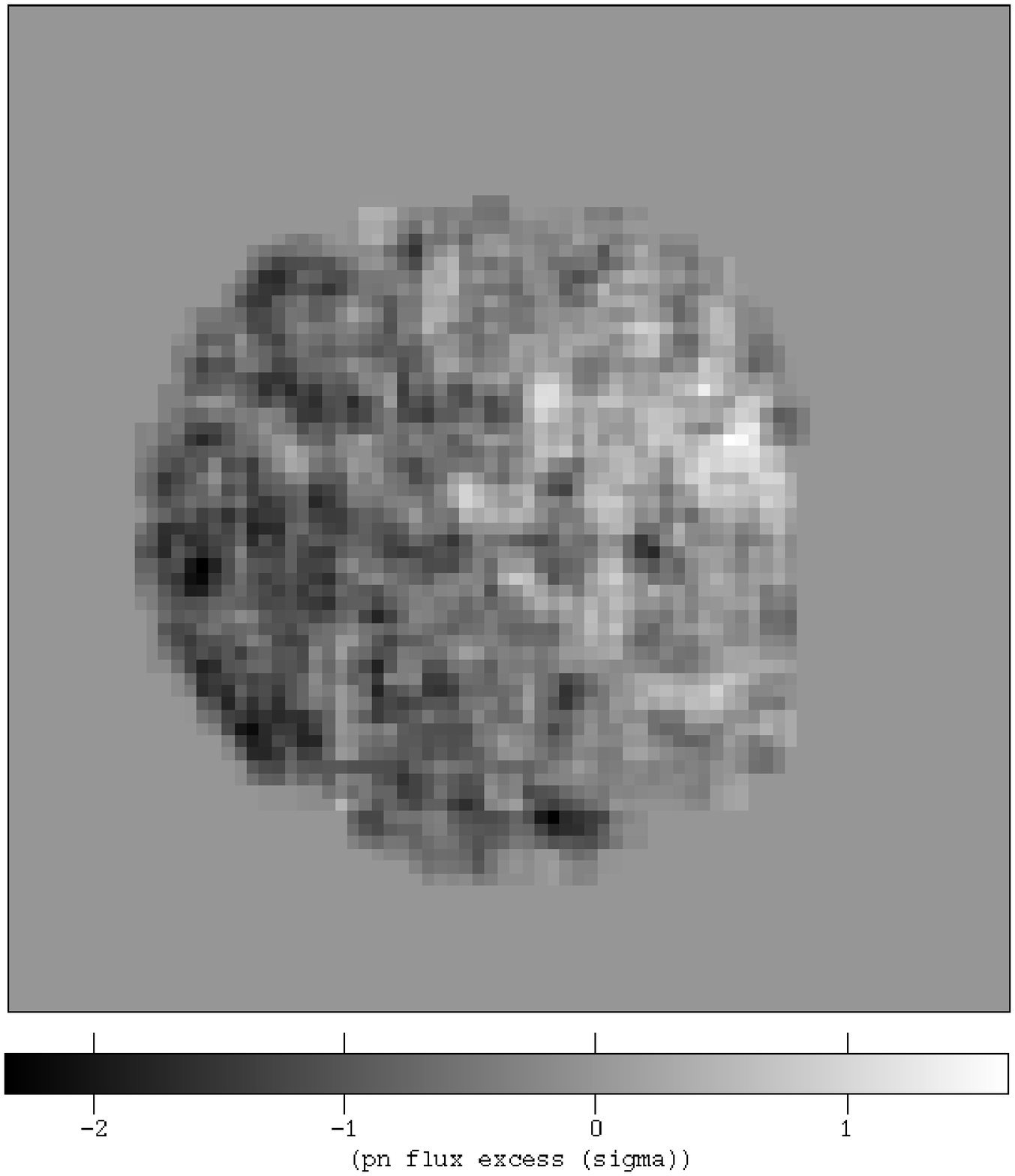}
\end{tabular}
\end{figure}

To highlight the effect visually, Figures~\ref{seren_image} and
\ref{seren_image2} show the excess flux per source 
detected in the 1XMM catalogue of EPIC serendipitous source detections
(plotted in units of sigma). The
images are displayed in the EPIC camera detector coordinates, and flux
determinations assume the nominal vignetting correction centred on the
reference pixel of the detector co-ordinate system (DETX, DETY in the
nomenclature of the XMM data analysis system). These figures show
that some low level discrepancy in the spatial variation in effective
area calibration must be present.

For XMM-Newton, direct measurement on the ground of the X-ray vignetting
function was prevented because nearly all X-ray beam measurements were
performed in a non-parallel beam. The installation of an X-ray stray-light
baffle in front of the mirrors, and the Reflection Grating Array (RGA) stack at
the mirror exit plane \cite{denHerder}, introduced potential complications
that were only measured in long wavelength, visible light at an EUV parallel
beam facility \cite{Tock}.

Although the measured {\em geometric} vignetting factor at longer wavelengths 
was comparable with predictions, it was necessary to use in-orbit data to 
confirm the X-ray energy dependence, and check that the geometric factor was 
maintained through the spacecraft assembly, integration, verification and 
launch campaigns.

\section{Telescope}\label{sect:mirrors}
The design of the XMM-Newton optics was originally driven by the requirement
to obtain the highest possible effective collecting area over a wide band of
energies up to 10 keV. The three, nominally identical, mirror systems
utilize a shallow grazing angle of $\sim$0.5$^{\circ}$ in order to provide
sufficient reflectivity at high energies. The effective area is increased by
nesting 58 mirror shells in each telescope to fill the front apertures as efficiently as
possible. Both the paraboloid and the hyperboloid sections of each shell
were replicated as a single piece from a single mandrel \cite{Chambure}.
Each telescope is complemented by an X-ray baffle, which minimizes X-ray
straylight from sources outside the field of view, when rays reach the focal
plane detectors by single reflection from the hyperbola.

The energy dependence of vignetting, which is superposed on any geometrical
component, becomes apparent at the critical angle for grazing incidence at
the off-axis angle of the target. There may be an increase in effective area
again at higher energies as a consequence of the fact that only the
innermost mirror shells provide substantial reflectivity. For a small
diameter shell, at high energies, the area increases initially with off-axis
angle: on one side of the mirror the parabola grazing angle is shallower
than for the on-axis geometry. The corresponding hyperbola graze angle is
then larger but because of the asymmetry of the reflectance vs. angle curve
the higher reflectivity on the parabola dominates the product of the
reflectances.

Each shell was individually aligned during assembly of the telescope
to a design accuracy of 10's arcseconds and glued into a mounting
``spider''. The whole mechanical assembly was provided with alignment
fiducials and optical alignment cubes and mirrors to ensure correct
placement during the various activities for on-ground calibration  and
assembly into the spacecraft. The nominal error budget allowed for
possible misalignment of the telescope axis of $\sim$30 arcseconds.

Two of the three telescopes were equipped with Reflection Grating
Arrays that also required alignment with the telescope axes. The three
telesope assemblies were mounted on a spacecraft mirror platform that
 contained star trackers providing an absolute reference for the
co-ordinate system of the spacecraft.

The focal plane detectors were aligned so that a reference
pixel was located on the nominal telescope optical axis. In the case
of the EPIC MOS cameras the reference was the central pixel of its
middle CCD, while for the PN camera a dead gap at the physical centre
of the camera, between CCDs, meant that the reference pixel was chosen
to be slightly offset to ensure that on-axis targets were not lost
onto the gap. The location of the RGA readout cameras were defined to
ensure that no expected bright emission lines in dispersed spectra
would fall on gaps between its CCD detectors. These reference pixels
were the origin of the DETX,DETY co-ordinate system of the XMM-Newton
Science Analysis System (SAS, \opencite{SAS}), in units of 0.05 arcseconds
per pixel.  

The physical alignment of the cameras was subject to possible error in
addition to uncertainty in locating the telescope axis, and the design
had to allow also for the possibility of 10's arcsecond relative
shifts due to the effects of launch loads, and eventual differential
shrinkage of the carbon-fibre optical bench tube due to 
water vapour out-gassing in orbit. One of the first in-orbit
calibration tasks was to determine the extent of any such shifts and
also determine the co-alignment to the spacecraft (star-tracker)
reference axis. For any observation a user could potentially request
any one of six (including the Optical Monitor) instruments to be the
prime, and for which a preferred detector location was defined to
avoid inter-CCD gaps. The mis-alignment for each of these ``boresight
axes'' had to be determined. Fortunately the initial mechanical
alignment was sufficiently good, and preserved into orbit so that a
single optimised boresight could be defined for all X-ray imaging
configurations, and one for RGA spectroscopic observations.
It should be highlighted that this complicated set of axis definitions
is a peculiar consequence of the multiple telescope configuration of
XMM-Newton compounded by the use of different instruments in the same
telescopes and between different telescopes.    
 \section{SuperNova Remnant Data} 
\subsection{Observation Configuration for SNR Targets}\label{sect:obsv}

The measurement of vignetting requires a compact, simple-spectrum,
non-variable source at locations off-axis, with which to compare the
inferred spectrum with that of the same object measured on-axis. True point
sources with reasonable brightness are precluded because the effects of
pile-up \cite{ballet} are severe, and furthermore vary with the changes in
off-axis Point Spread Function (PSF), as well as with the count rate
reduction due to the vignetting itself.

\begin{figure}
\caption{Schematic drawing of the CCD orientations in the 3 co-aligned EPIC cameras. }  \label{gaps}
  \begin{center}
\begin{tabular}{c}
  \includegraphics[bb = 30 169 565 680, clip,scale=0.65]{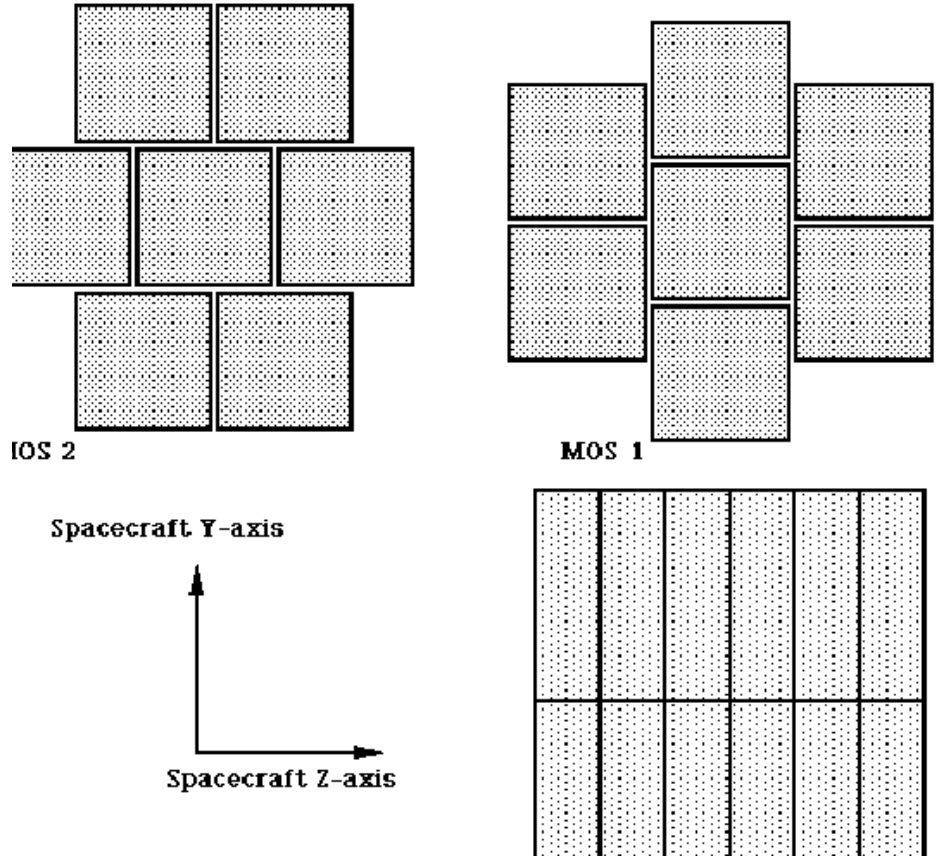}
\end{tabular}
  \end{center}

\end{figure}

Extended objects require a complex ray-tracing and PSF-folding to account
properly for the vignetting component. While a number of viable targets were
selected for the in-orbit calibration, we have concentrated on G21.5-09
\cite{warwick} and 3C58 \cite{3c58} for this work. The initial choice of
pointing locations was complicated by the need to ensure that no significant
portion of the remnants fell near CCD gaps. Given the orthogonal orientation
of the two MOS cameras, together with the totally different
gap patterns in the pn (Figure~\ref{gaps}), this severely constrained
the orientation available, and an angle $\sim$7 degrees rotated from the
nominal detector axes, and a field angle of 10 arcminutes were chosen for
the initial measurements of G21.5-09. 

\begin{figure}
\caption{Merged pn image of the 5 major pointings made on G21.5-09 SNR. White gaps are physical gaps between CCDs or noisy columns}  \label{G21image} 
 \begin{center}
\begin{tabular}{c}
    \includegraphics[bb= 197 324 397 519, clip]{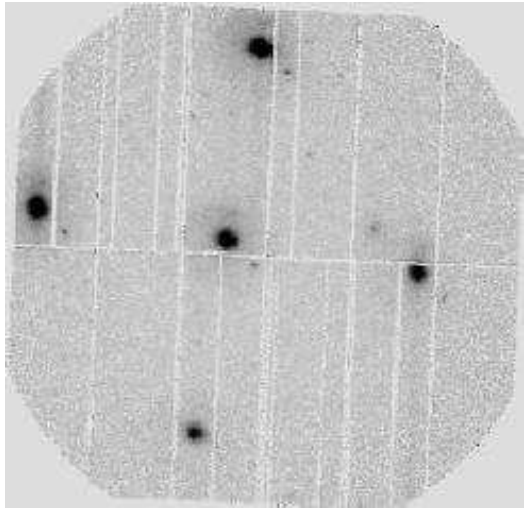}
\end{tabular}
 \end{center}

\end{figure}

As a consequence of the grating array angles and blocking fraction, the
vignetting in the MOS cameras is expected to be a strong function of
azimuthal angle, so four locations were scheduled for G21.5-09 
to sample the extreme ranges of RGA blocking (see Figure~\ref{G21image}).

\begin{figure}
\caption{Merged image of the major pointings made on SNR 3C58}\label{3c58images} 
 \begin{center}
\begin{tabular}{c}
    \includegraphics[bb = 182 275 429 516, clip]{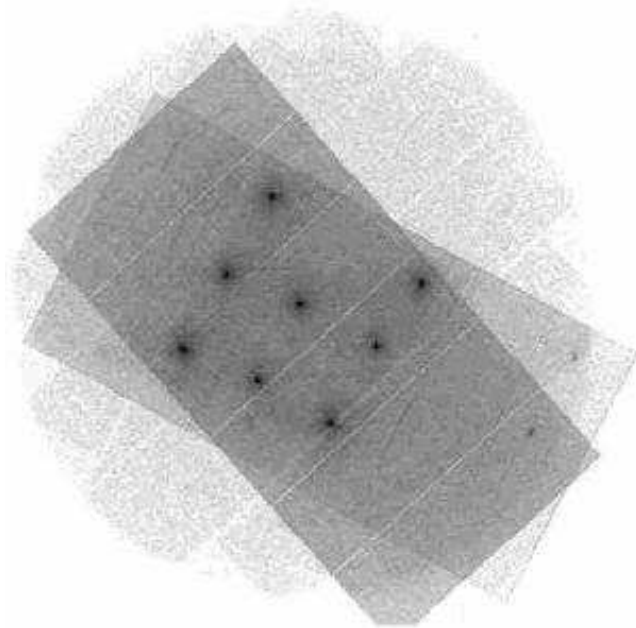}
\end{tabular}
 \end{center}

\end{figure}

SNR G21.5-09 is a hard, bright source ideal for determining the energy
dependent vignetting. A set of observations of the softer, somewhat fainter,
SNR 3C58 were subsequently made to supplement the earlier observations and
better sample the azimuthal dependency of the MOS camera vignetting.
Measurements of 3C58 were scheduled to fall within the central CCD of the
MOS cameras, to minimize the possible effect of varying quantum efficiency
over the detector. Again the limitations of chip gaps cause a compromise in
the actual rotations employed, such that the ensemble of pointings is as
shown in Figure~\ref{3c58images}.
 
Table \ref{t:obs} summarises the requested pointings and observation
details for these data sets

\begin{kaprotate}
\begin{table}[ht]
\caption{Summary data for the SNR observations used in
  the analysis} \label{t:obs} 
\begin{tabular}{ccclcc} \hline\hline
Target & Revolution & Observation ID& Date & RA 2000 & DEC 2000\\ \hline\hline
G21.5-09  & 60 & 0122700101 & 2000-04-07T12:35:28 & 18:33:33 & -10:34:01\\
G21.5-09 S& 61 & 0122700201 & 2000-04-09T12:22:17 & 18:33:40 & -10:44:18\\
G21.5-09 W& 62 & 0122700301 & 2000-04-11T12:25:38 & 18:32:52 & -10:35:47\\
G21.5-09 N& 64 & 0122700401 & 2000-04-15T12:25:52 & 18:33:26 & -10:24:02\\
G21.5-09 E& 65 & 0122700501 & 2000-04-17T12:13:09 & 18:34:14 & -10:32:32\\
G21.5-09 NE&244 & 0122701001 & 2001-04-09T17:06:14 & 18:34:23 & -10:30:40\\\hline
3C58    &506 & 0153752101 & 2002-09-13T04:22:11 & 02:05:38 & +64:49:40\\
3C58 W  &505 & 0153752201 & 2002-09-11T04:29:35 & 02:04:43 & +64:51:13\\ 
3C58 SW &505 & 0153751801 & 2002-09-11T13:09:58 & 02:05:03 & +64:47:38\\
3C58 S  &505 & 0153752501 & 2002-09-11T20:03:41 & 02:05:23 & +64:43:52\\
3C58 E  &505 & 0153752401 & 2002-09-12T03:20:44 & 02:06:32 & +64:48:06\\
3C58 NW &506 & 0153751701 & 2002-09-13T10:57:39 & 02:05:18 & +64:53:23\\
3C58 NE &506 & 0153751901 & 2002-09-13T17:51:22 & 02:06:13 & +64:51:41\\
3C58 N  &506 & 0153752001 & 2002-09-13T23:45:05 & 02:05:52 & +64:55:27\\ \hline
\end{tabular}
\end{table}
\end{kaprotate}

\subsection{Analysis of SNR Pointings}\label{sec:anal}
Spatial regions of interest were defined around the centroid of each
SNR, and fixed in sky co-ordinates for all observations, ensuring that
the region was large enough compared with the PSF, but small enough to
avoid a CCD gap in any single observation. Background regions of
similar size were selected. Concerns about enhanced particle
background in some of the observations led us to choose regions at
similar field angle and azimuth to ensure representative conditions,
while strict selection of low background count rate intervals
minimised any systematic effects of any possible incomplete background
subtraction.   The analysis proceeded by determining the number of
background-subtracted photon counts per energy bin at various {\em pseudo}
on- and off-axis locations of G21.5-09 or 3C58. The relative vignetting
between the corresponding locations were determined according to the count
rate variations. The energy bins' widths were varied semi-logarithmically to
maintain reasonable signal:noise per bin. 
\clearpage

Unless there were gross misalignments of shells within a telescope,
the vignetting in the EPIC PN telescope would be radially symmetric
about its optical axis. True azimuthal variations are expected in the
MOS telescopes as a consequence of differential shadowing according to
the angles of the RGA grating plates. Verification of the predictions
for these MOS azimuthal variations in vignetting
were undermined by unexpected and significant variations ($\sim$10\%) in
relative vignetting measured in the pn camera, from azimuth to azimuth.
This was attributed initially to a combination of incomplete background
correction and to discrepancies in the exposure dead-time calculations
influenced by the higher than nominal background. Eventually it was realised
that these relative variations were correlated with camera
orientations.

It was recalled that unresolved discrepancies between mirror optical alignment
cube axes and inferred telescope axes measured at the EUV test facility
\cite{csl} had occurred. At the time these orientation discrepancies
were claimed to be irreproducible to $\sim$20 arcsecond level, but
were also seen in similar magnitude and direction in the Panter X-ray
test facility calibration of maximum throughput orientation
\cite{panter}. The variation was in addition to any {\em fixed and  systematic}
offset between the mechanical and optical telescope module axes
(designed to be less than 30 arcseconds).   For the PN camera, we
therefore simply vary the location of assumed telescope optical axis
and recompute the expected vignetting function at each observed
location. The axis origin is defined where the difference  in low
energy  vignetting for all measured points is minimised. 

 Once reasonable agreement for the PN data was obtained, we proceeded
 to treat the MOS cameras in a similar manner, except in this case the
 azimuthal effect of RGA shadowing partially mimics a potential axis
 shift. We therefore {\em assumed} the nominal RGA performance in the
 calculation. In principal the two effects can not be distinguished,
 except: \begin{itemize}
\item A gross error in RGA shadowing angles would have been detected
  in the dispersion relation and/or effective area of that instrument
  (not the case)
\item For most science analysis we just need to have an empirical verification of
  the MOS  vignetting function, whatever the cause of the azimuthal
  changes
\end{itemize}
 
\subsubsection{Vignetting Results}\label{sec:results}

\begin{figure}
\caption{(a - upper) Relative (i.e. compared with
    nominal on-axis location) vignetting of the pn telescope for an
    off-axis angle of 11.3 arcminutes from the nominal boresight
    location. (b - lower) {\em Relative} vignetting of the MOS1
    telescope for an off-axis angle of 10.4 arcminutes, compared with
    the nominal boresight location. The theoretical prediction for energy
dependent vignetting is shown as a solid line in each case. This is
    the leftmost observation depicted in Fig~\ref{G21image}}\label{PNR65} 

  \begin{center}
    \includegraphics[scale=0.58]{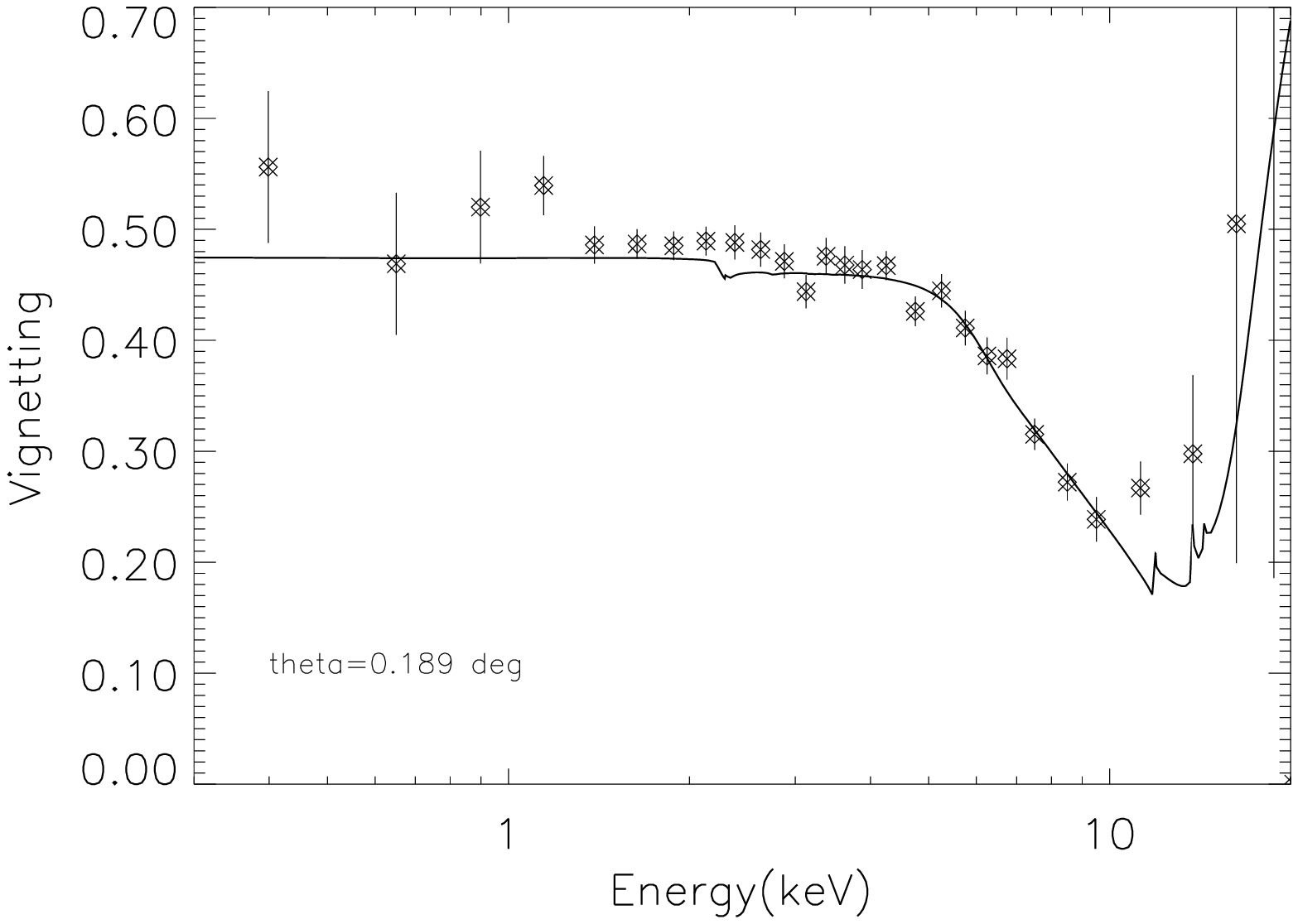}
    \includegraphics[scale=0.58]{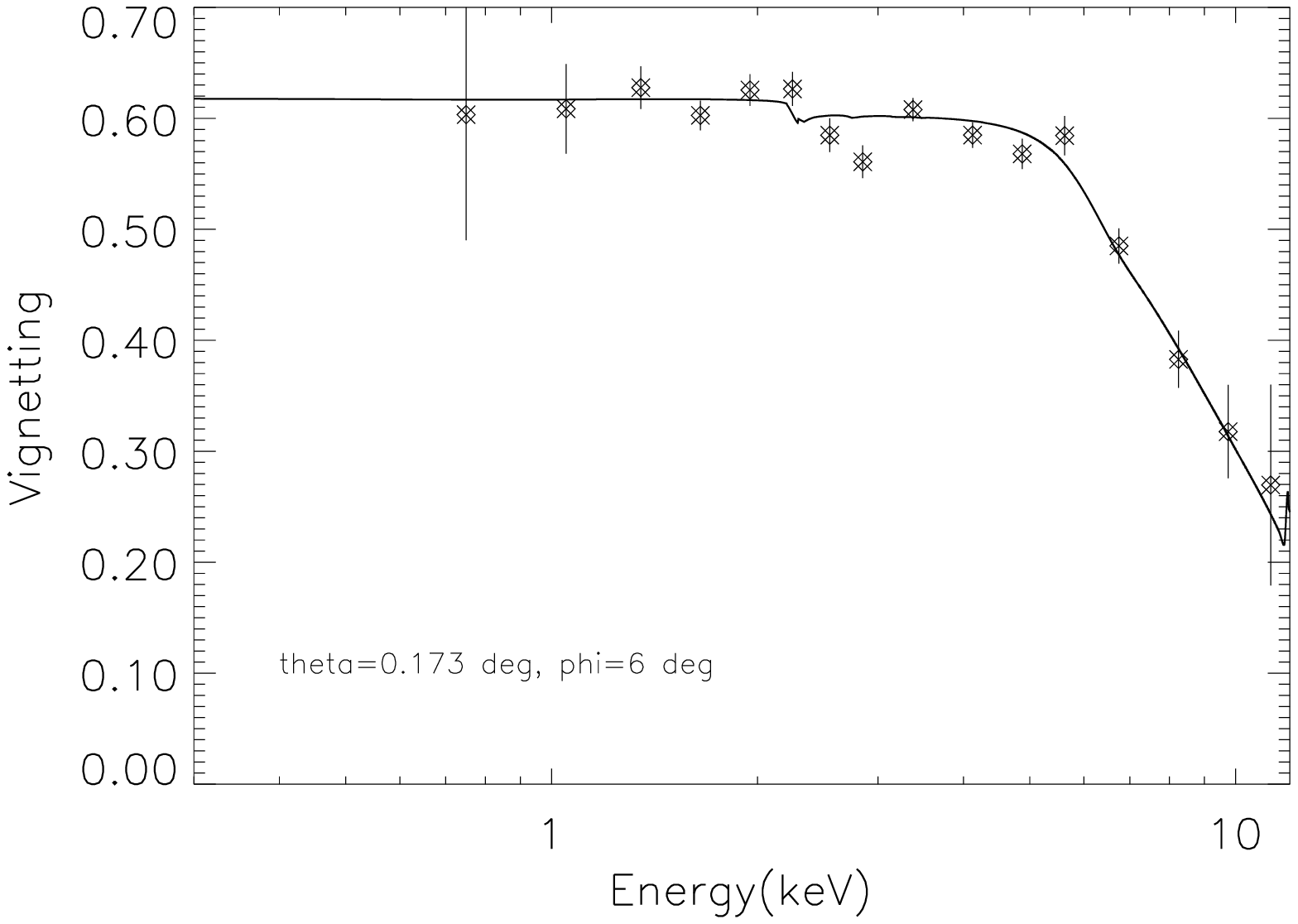}
  \end{center}

\end{figure}

A single azimuth vignetting measurement for the pn camera, in the lowest
background exposure, is shown in Fig.~\ref{PNR65}a.  A subsequent
calibration observation in circa April 2001 at larger off-axis angle allowed
some measure of sensitivity to the change in $\theta$ (off-axis angle), and confirms the
validity of the model.

A comparable vignetting measurement for the MOS cameras is shown in
Figure~\ref{PNR65}b. Due to the lower effective area of the MOS cameras,
the S:N is lower than for the pn camera, and the energy scale is binned more
coarsely. It was found, as with the pn camera, that there was a potential telescope axis
misalignment. However records of the tests in ground facility were
less clear than for those of the pn, because the installation of the RGA had blocked the access to the
mirror alignment lens for most tests. Relying purely on inferred alignment
of the axis based on the vignetting itself undermines the goal of directly
measuring the effect of RGA azimuthal blocking factor.

\subsubsection{Energy Dependence}
\begin{figure}
\caption{{\em Relative} vignetting of the pn telescope after 
averaging all azimuths around 10.3 arcminutes off-axis. The energy dependence is in 
good agreement (solid line)}  \label{PNAV}
  \begin{center}
\begin{tabular}{c}
    \includegraphics[scale=0.58]{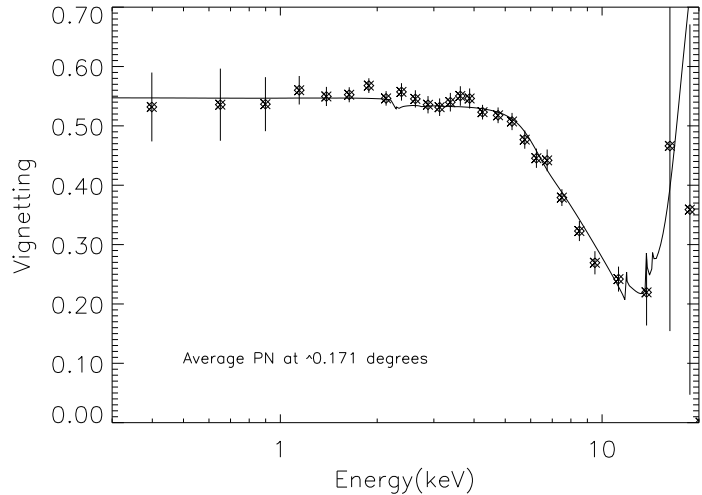}
\end{tabular}
  \end{center}

\end{figure}
The pn data sets were relatively close in off-axis angle and should have no
intrinsic azimuthal dependence. We should be able to average the 4
separate locations of G21.5-09 to check the predicted energy 
dependence is correctly reproduced. This is shown in Fig.~\ref{PNAV}.

\begin{figure}
  \begin{center}
   \includegraphics[scale=0.58]{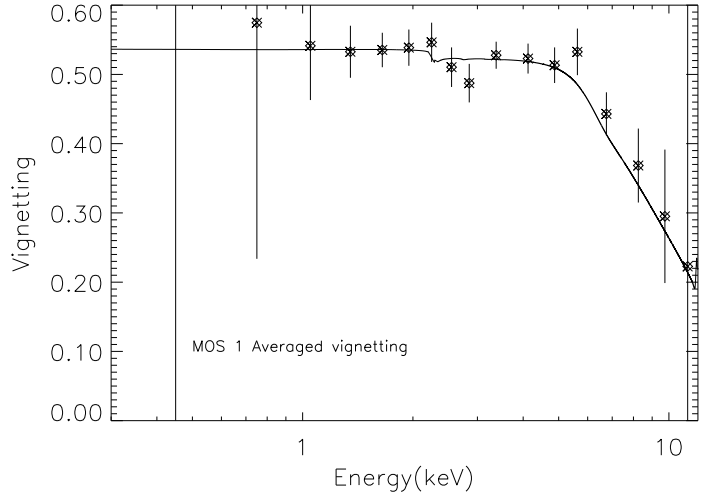}
  \end{center}
\caption{{\em Relative} vignetting of the MOS1 telescope 
after averaging all azimuths around 10.3 arcminutes off-axis. The energy
dependence is in good agreement}\label{mos1vigav} 
\end{figure}

\begin{figure}
  \begin{center}
\includegraphics[scale=0.58]{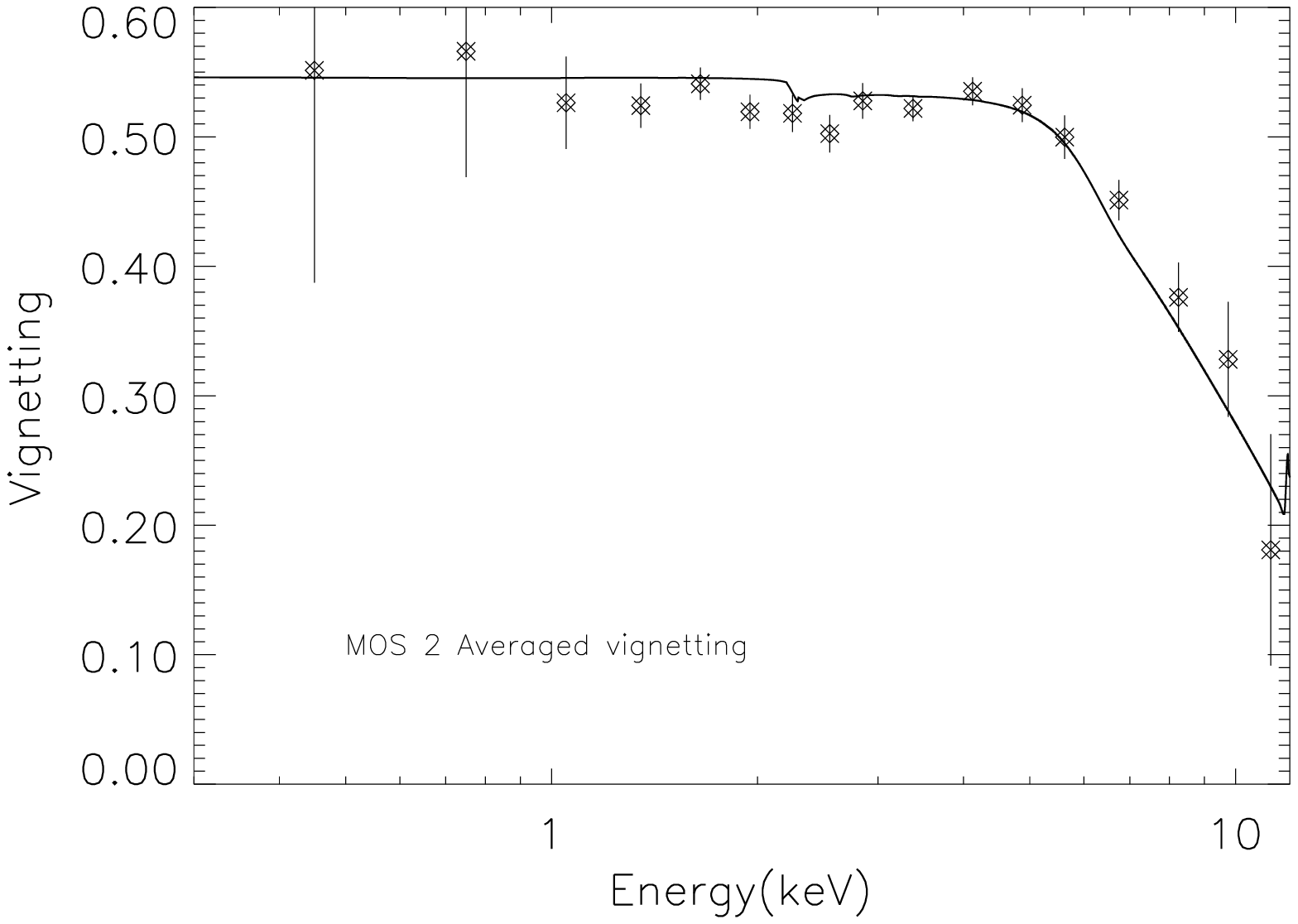}
   \end{center}
\caption{{\em Relative} vignetting of the MOS2 telescope 
after averaging all azimuths around 10.3 arcminutes off-axis. The energy
dependence is in good agreement}\label{mos2vigav} 
\end{figure}

For the MOS data, repeating the exercise is not really valid, given the large
variation in RGA blocking with azimuth. However to discern if the placement
of RGA gratings and ribs upsets the energy-dependent filter properties via.
differential shadowing of some sub-sets of shells, we nevertheless form the
same average response in the 2 MOS cases. There seem to be no significant
energy-dependent discrepancies (see Figures~\ref{mos1vigav} and \ref{mos2vigav}).


\subsubsection{Location of optical-axes}

\begin{figure}
  \begin{center}
   \includegraphics[scale=0.42,angle=270]{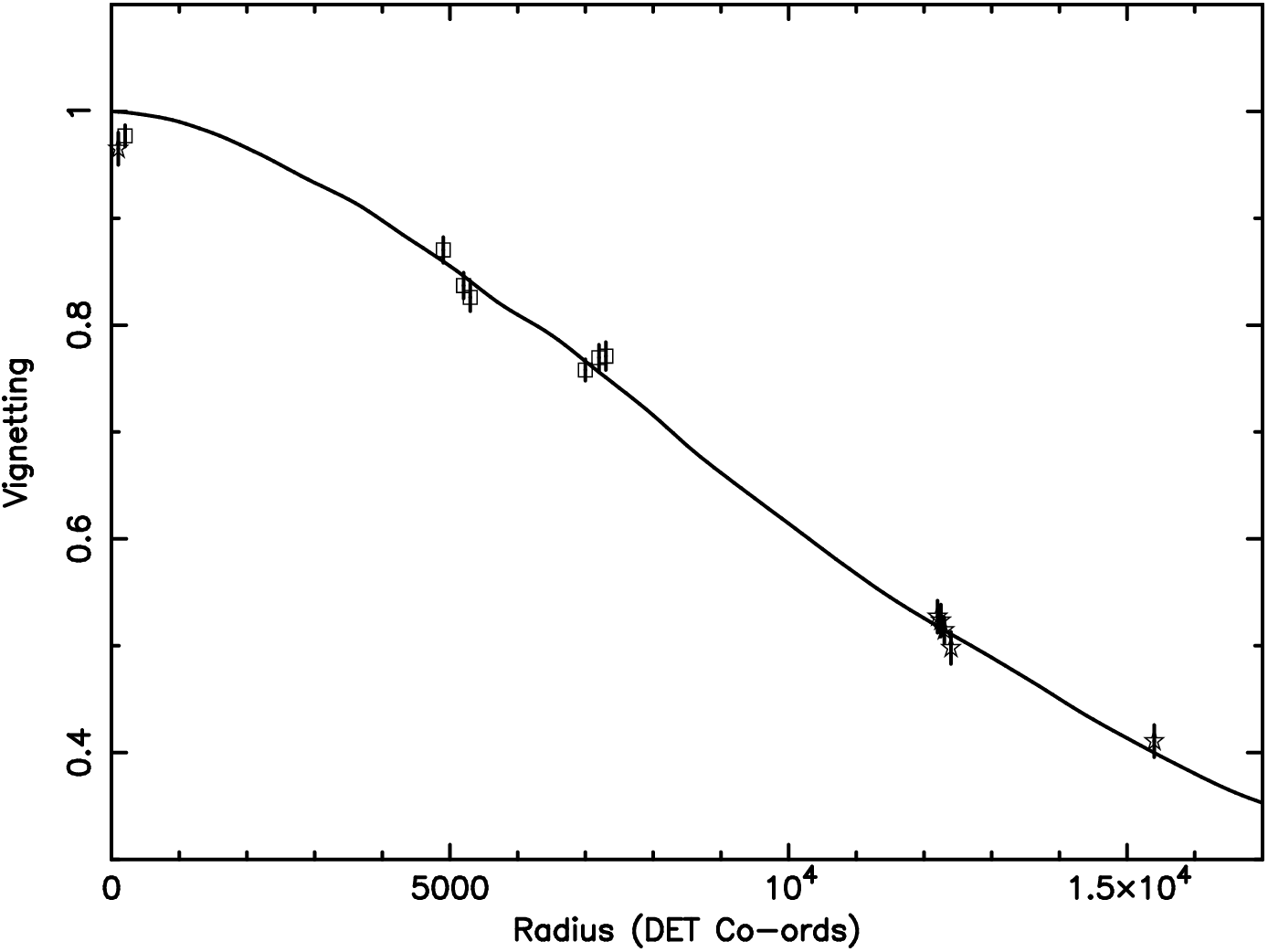}
  \end{center}
\caption{Comparison of G21.5-09 (stars) and 3C58 (squares)
MOS-1 data with the expected mirror vignetting, centered at DETX=200, DETY=-50.
The data have been corrected for RGA blocking.}\label{mos1theta} 
\end{figure}

\begin{figure}
  \begin{center}
   \includegraphics[scale=0.42,angle=270]{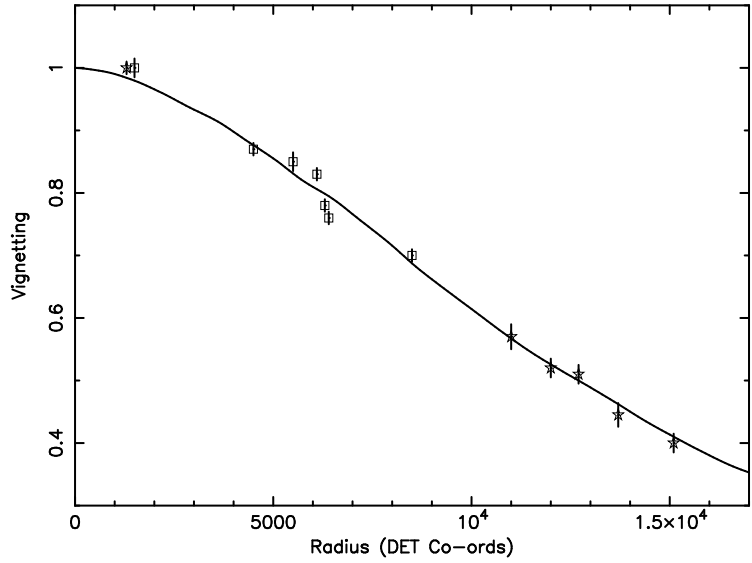}
  \end{center}
\caption{Comparison of G21.5-09 (stars) and 3C58 (squares)
MOS-2 data with the expected mirror vignetting, centered at DETX=340, DETY=-1300.
 The data have been corrected for RGA blocking.}\label{mos2theta} 
\end{figure}

\begin{figure}
  \begin{center}
   \includegraphics[scale=0.42,angle=270]{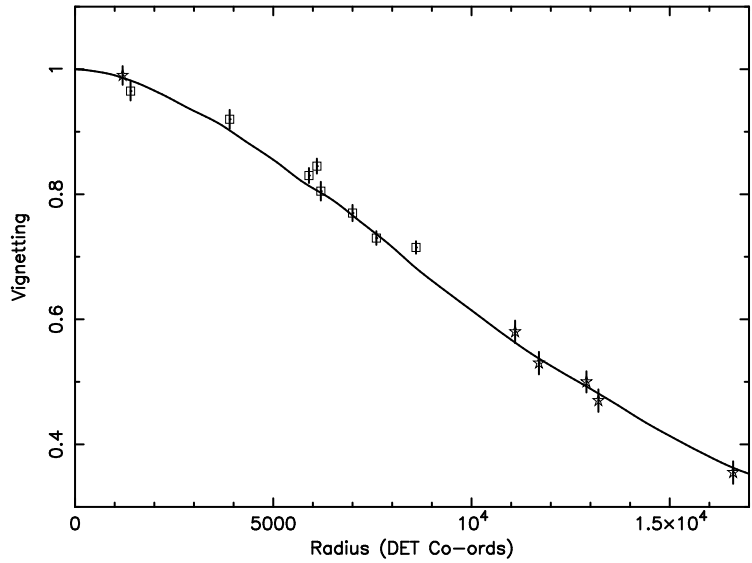}
  \end{center}
\caption{ Comparison of G21.5-09 (stars) and 3C58 (squares)
pn data with the expected mirror vignetting, centered at DETX=1300, DETY=450.}\label{pntheta}
\end{figure}

It can be seen that the energy dependence of vignetting is almost negligible
in these data sets up to $\sim$4keV. Initially therefore, spectral
parameters were independently found for each SNR by a joint fit of an
absorbed power-law to all observations with energy range limited to E$<$
4keV and varying normalizations per observation, in order to reduce
statistical effects of spectral determination.  The best-fit spectral model
was then applied individually to each observation to find the relative
normalizations, and the energy dependence for E$>$4keV.  After
renormalization the combined data of G21.5-09 and 3C58 were used to locate
the optical axis of each camera using a minimization technique.  A good fit
to the predicted low energy vignetting, as a function of off-axis and
azimuthal angle, was achieved by applying a small shift in the optical axis
position for MOS-1 (Fig.~\ref{mos1theta}). A larger offset, of the order
of 1 arcminute, was found to be necessary to obtain a good fit for the MOS-2
and pn telescopes (Figs.~\ref{mos2theta},\ref{pntheta}).

\section{Alternative Measurements}
In order to provide corroborating evidence for these axis shifts, different 
measures were proposed. 

\subsection{Source elongation}\label{sec:elong}
\begin{figure}
  \begin{center}
   \begin{tabular}{c}
   \includegraphics[scale=0.4,angle=270]{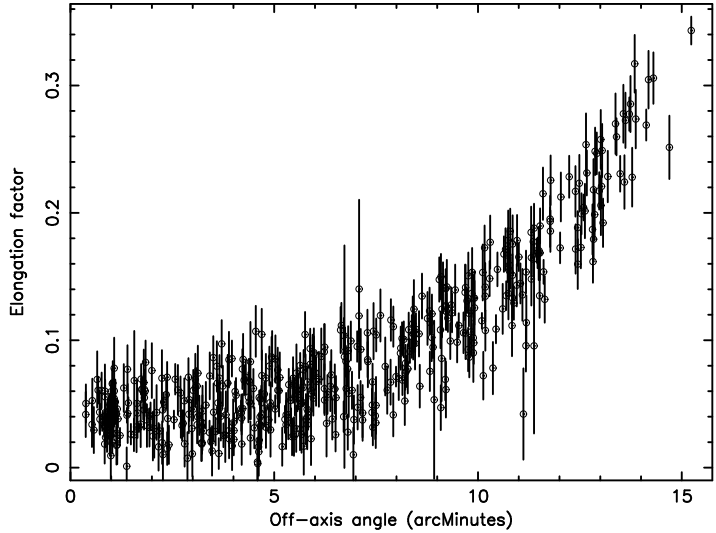}
\end{tabular}
  \end{center} 
\caption{ Relative elongation of point sources in the pn
  camera as a function of off-axis angle}  \label{elong}
\end{figure}

The PSF broadens with off-axis angle, especially in a direction perpendicular
to a vector connecting the source and optical axis positions. A 
plot of source PSF elongation versus off-axis angle should therefore be 
symmetrical
about the optical axis. We selected a large sample of serendipitously 
detected sources, after removal of non-point-like objects, from the 1XMM catalogue, for each camera. Figure~\ref{elong} shows the variation of this
elongation with off-axis angle for the pn sample. The centroid of the
distribution was found by minimizing the function:
\begin{center}

        $   E = A + B\theta^{2} + C\theta^{4}  $

\end{center}
where $E$ is the measured elongation, $\theta$ the off-axis
angle measured from the telescope axis and
$A$,$B$ and $C$ are coefficients.

The elongation $E$ was defined as follows:\begin{itemize}
\item the source image out to a cutoff radius of 20 arcsec was resampled in source-centric polar coordinates r and $\phi$;
\item the resulting image was multiplied by r to preserve scaling, then
Fourier-transformed in the $\phi$ coordinate;  
\item the ratio between the magnitudes of the 0$^{th}$ and 2$^{nd}$
  Fourier coefficients was taken as the elongation.
\end{itemize}
 
The best fit centroids reveal a qualitatively similar
axis shift to that measured with G21.5-09 and 3C58. The measured
values are: PN (DETX,DETY =  1140,340),  MOS1 (DETX,DETY = -320,+540)
and MOS2 (DETX,DETY = -340,-1700). 

A small discrepancy was noted for the MOS cameras, again suspected to be due
to the effects of the RGA assembly. A raytrace for a nominal XMM telescope
was made, in which multiple sources were traced through to the focal plane,
and then their elongations in focal spot determined as a function of
position in the focus, by simple Gaussian fitting. For an unobscured
telescope, the minimum elongation occurred at the centre of the focal
plane. However for a telescope fitted with a nominal RGA structure, the
minimum elongation occurred off-axis, in a direction parallel with the
dispersion axis. It is believed this can be attributed to shadowing of rays
that are intercepted by the grating plates, restricting the scattering (and
hence elongation) in a direction parallel with the dispersion direction.  As
the RGA plates are angled to the telescope axis, the location of elongation
minimum is hence offset. The ray trace estimate of this offset is about 22
$\pm$1 arcsec.
\subsection{Diffuse Background}
In \cite{back}, the compilation of a set of data from high galactic fields
was described. After exclusion of individual point sources, this field
represents the average properties of the diffuse Cosmic X-Ray
Background (CXB). Although
this background represents the superposition of many unresolved faint
sources, the averaging over several fields with XMM-Newton ensures any
``cosmic variance'' is minimized, and the artificial field should be very
uniform. In such a case the centroid of surface brightness distribution
should map to the optical axis of maximum throughput.

This is difficult to interpret in the case of the MOS cameras in particular
because the apparent vignetting varies in a direction parallel to the
grating array dispersion direction. The differential blocking of the grating
must therefore be unfolded from any potential  telescope axis tilt
effects.

\begin{figure}[ht]
  \begin{center}
   \begin{tabular}{c}
   \includegraphics[scale=0.55]{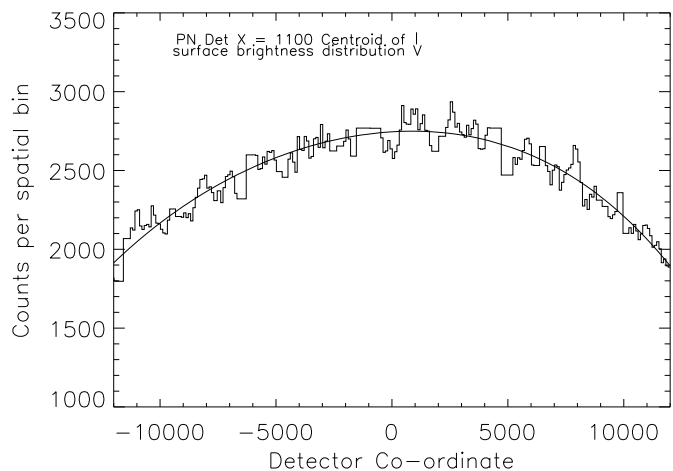}
\end{tabular}
  \end{center}
\caption{Surface brightness (0.5--4 keV) in 
spatial bins of the merged image of several high latitude background
fields. The solid curve is the best polynomial fit. The lack of
symmetry is attributed to the offset of the circular aperture of the
camera from the centre of the telescope}\label{SB} 
\end{figure}

The data set from high galactic latitudes was binned into images in the
energy band 0.5-4keV for each camera. The surface brightness distribution
was projected in detector X, and Y co-ordinates and was corrected for
exposure effects (bad pixels, CCD gaps, CCD dead times etc.). Next the
predicted amount of RGA blocking as a function of $\theta , \phi$ was
calculated by raytrace, and divided into the surface brightness
profiles. The centroid of brightness was then calculated by a polynomial
fit. See as an example Fig.~\ref{SB}.

The main weakness of this approach is that the RGA blocking is assumed
to be correctly modeled, and not left as a free parameter. The
justification is supported by the nominal performance of the RGA
dispersion properties as measured in-orbit by the spectrometer instrument.
\subsection{Coma Cluster}

\begin{figure}[ht]
  \begin{center}
   \begin{tabular}{c}
   \includegraphics[scale=0.5]{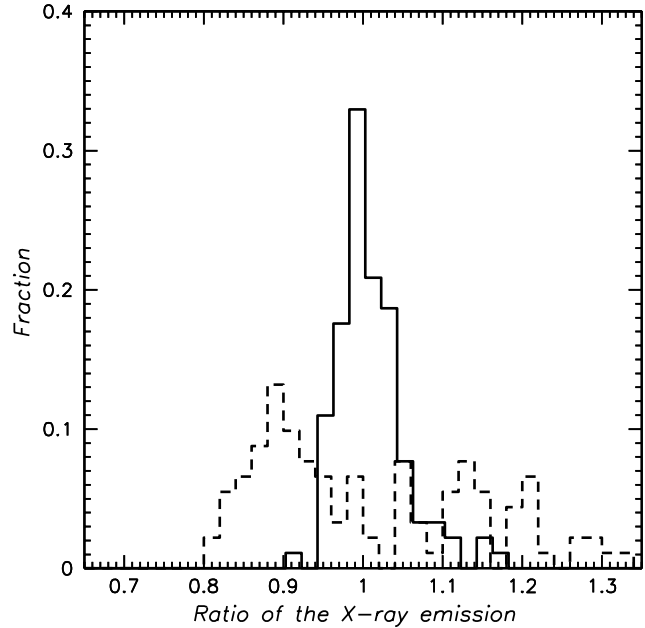}
\end{tabular}
  \end{center}
\caption{ Ratio of surface brightness (0.5--2 keV) in 
spatial bins of the Coma Cluster, for two different camera
orientations. {\em Dotted line} - vignetting corrected with initial
calibration. {\em Solid line} - vignetting corrected according to improved
estimate of telescope axis}\label{coma}
\end{figure}

The Coma cluster of galaxies is one of the brightest diffuse X-ray objects on
the sky, filling the field of view of XMM-Newton detectors. An additional
possibility to calibrate the vignetting of the mirror system of Epic pn is
provided by performance of a special calibration observation of the Coma
centre in addition to the existing set, described in Briel et
al. (2001). The idea is to use the same region on the sky but with
a position angle of $\sim120$ degrees between observations.  In this
case a vignetting miscalibration producing an
under-corrected part of the emission in the first observation will be
compared to over-corrected part in the second.
In case of the shift of telescope axis, such deviations occur symmetrically.
so that a ratio map of the two images has one side of the image being
larger and the other smaller than unity.

Additional fluctuations on the ratio map are introduced unless OOTE
(Out-Of-Time Events) subtraction is performed. This is a specific
feature of the pn camera (compared to MOS where OOTE are small in
fraction), where events are accumulated in the CCD while the previous
image frame is still being read out. The OOTE may be removed
statistically, using off-line products generated by XMMSAS {\it
  epchain} program for every observation. In this particular instance
the importance of
OOTE is caused by a change in the read-out mode for pn in the second
observation to full frame from extended full frame mode, that was
employed for the other XMM  Coma observations, such 
that this fraction of smeared events changed from 0.023 to 0.063.

We choose the $0.5-2$ keV band for the primary analysis, while
energy-dependent effects where checked using the $0.25-0.5$ keV and $5-7.9$
keV bands, considering previous data from mirror test results from the
PANTER facility as well
as the presence of strong background lines $>7.9$ keV.

In order to achieve good statistics a binning of $32\times32$ of original
pn pixels was employed ($128\times128$ for MOS). Border pixels in this
binning can have much lower fraction of valid pixels and were excluded from the
analysis. The best telescope position was chosen to achieve the lowest
scatter among about a hundred independent points of the image ratio
map, for both pn and each MOS 
camera. The effect of the change in the position of the telescope axis was emulated in
the calibration version of the exposure map, where by changing the
vignetting we still retain proper position of the detector chips.

%
%

In deatil the procedure comprised two main parts: preparation of the dataset and a
loop of calculations repeated with varied input position of the telescope
axis to minimize the spread in the dataset points.

The data preparation part comprised:
\begin{itemize}
\item screening both observations for background flares, as described in Briel
  et al. (2001).
\item for both observations extracting source and OOTE images  in the
  $0.5-2$ keV band 
\item Correcting both images for OOTE
\item subtracting the instrumental background, using Filter Wheel Closed
background accumulation (for description of this background dataset see
e.g. \cite{back})
\item Translate the image for one observation to the reference frame of the
  other 
\item Calculate the mask file where both observations have sufficient data. Only
$FLAG=0$ events are considered (large detector gaps). Also only the pn event
types with $PATTERN<5$ are considered.
\item Discard image pixels absent in at least one observation, using the
cross-correlation of the mask files for both observations. The resulting
mask file is retained to be later applied to experimental exposure maps.
\item bin the image to achieve good statistics.
\end{itemize}

The calibration loop comprised:
\begin{itemize}
\item For input parameters of the test telescope axis position create
exposure maps for both observations
\item Normalize the exposure map to correct for loss of the flux due to
subtraction of OOTE, by multiplying the maps by $(1-f_{OOTE})$, 0.937 and
0.977 for full and extended full frames, respectively
\item Translate the exposure map for one observation to the
  reference frame of the other
\item Discard exposure pixels absent in at least one observation, using the
  pre-calculated mask file
\item Bin the exposure maps in accordance to binning of the image
\item De-vignette the images and produce their ratio
\item estimate the dispersion around the mean. The position of the telescope
  axis is searched to minimize this dispersion.
\end{itemize}

Uncertainty in the parameter estimate is calculated using 90\% confidence
level estimate for the $\chi^2$ method. However the systematic
uncertainty seems to drive the error, and is related to an astrometric
uncertainty for the position of each image.

Epic pn data comparison shows a very small dispersion at a telescope axis
position $DETX=1243\pm30$ pixels $DETY=402\pm30)$ pixels in detector
coordinates. The one-dimensional scatter of pixels around the mean is
plotted in Figure~\ref{coma}. Note that the mean value of 1.000 is not
enforced and is an additional argument in favor of the method.
The original calibration introduced a 14\% r.m.s. scatter in the surface
brightness data, while the proposed revised calibration decreases the
r.m.s. to 3\%, comparable with the statistical noise. This result is
consistent with the large r.m.s. scatter found in pn-MOS1 comparison of
serendipitous sources (see Figures 1 and 2 and \opencite{sax}).

Table \ref{t:coma-pn} illustrates the quality of the calibration.

\begin{table}[ht]
\caption{ Residual dispersion in the pn vignetting
  calibration. } \label{t:coma-pn}
\begin{tabular}{ccc} \hline\hline
Radius        &       \multicolumn{2}{c}{Dispersion, \% }\\
arcmin  &   best fit &        at 2 sigma deviation\\
\hline
 8--11    &   3.712   &      3.826\\
 4--8     &   2.376   &      2.422\\
 0--4     &   2.186    &     2.199\\
 \hline
\end{tabular}
\end{table}

A change in the position twice the quoted error bar, presented in Table~\ref{t:coma-pn}
above produces noticeable changes on the image being dimming and brightening
of the opposite sides. Variation in the background level can introduce
additional 1\% errors for furthermost from on-axis pixels.

The adopted error on the position of the telescope axis includes various
systematics e.g. caused by the small misalignment between the two Coma pointings and
changes in the instrumental background level, while the formal statistical
error is smaller, 10 pixels. However, the precision at which the position of
the telescope axis is achieved exceeds by far the typical resolution at
which e.g. the exposure maps are calculated (200 DETX/DETY pixels or
10 arcseconds).

\begin{figure*}[ht]
\includegraphics[bb = 142 27 473 767,scale=.45,angle=270]{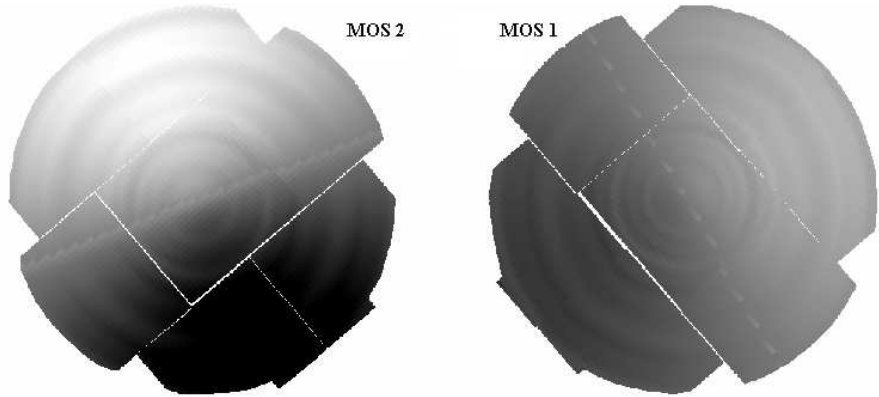}
\caption{Ratio of the new and old exposure maps for MOS 1 and 2, showing
  the amplitude and direction of the change. Fine structure on these images
  (a line and annuli) is an artifact produced by finite precision at which
  the exposure maps are generated. Vignetting is azimuthally
  symmetric, while the line-like feature of CCD gaps are parallel
  to the direction of equal RGA transmission, and not this vignetting shift.
}\label{f:mos_2d}
\end{figure*}

\subsection{Calibrating MOS vignetting using the Coma observations}

Calibration of the MOS vignetting using the Coma cluster has been first done
by cross-calibration with pn and later repeated using a self-calibration
method, as has been described above for pn. Table \ref{t:coma-m} lists three
independent measurements of the positions of the telescope axis obtained
using both Coma MOS observations, 2001 and 2000 denoting the MOS-pn
comparison, using the year of MOS observation to describe the observational
dataset, the 'average' of the two and 2000 vs 2001 'self' calibration of
MOS.

%
%

\begin{table}[ht]
\caption{Positions of the MOS telescope axis from the analysis
  of Coma cluster. Units are internal camera values of 0.05
  arcsecond.} \label{t:coma-m} 
\begin{tabular}{cccc} \hline\hline
Dataset & DETX & DETY & Dispersion, \%\\
\hline
\multicolumn{3}{c}{MOS2}\\
2001 & $653\pm60$ & $-1260\pm30$ & 4.50\\
2000 & $440\pm60$ & $-1250\pm30$ & 4.65\\
average & $550\pm60$ & $-1255\pm30$ &\\
self    &  $492\pm60$ & $-1256\pm30$ & 4.73\\ \hline
\multicolumn{3}{c}{MOS1}\\
2001 &   $136\pm40$     & $-134\pm70$ & 4.56\\ 
2000 &   $87\pm40$     & $-281\pm70 $ & 4.76\\
average & $110\pm40$     & $-200\pm70$ &\\
self    &  $159\pm40$ & $-303\pm70$ & 4.68\\
\hline
\end{tabular}

\end{table}

The difference in the normalization between 2000 and 2001 MOS1 observations
amounts to 6\%, and is due to differences in the best-fit positions. The
self-calibration finds the mean flux ratio between two observations of
1.005, which is acceptable. The similarity of the results of cross and self
calibration of MOS is reassuring and demonstrates negligible effect of
possible chip-to-chip sensitivity variations for MOS. However, the
dispersion achieved in the best-fit position is worse, compared to pn
results and could partly be caused by lack of fidelity in the description of the
RGA shadowing in the current version of SAS.


Fig.\ref{f:mos_2d} shows the magnitude of the proposed changes in the
vignetting for MOS 1 and 2 cameras. Only a small change is proposed for MOS
1, while vignetting for MOS 2 is substantially revised. The effect of
RGA-vignetting degeneracy reduces the sensitivity of the method to the
absolute value of the telescope shift, yet the direction of the shift is
rather well determined. Since the problem originates in the low contrast in
the exposure map, caused by these changes, it implies little importance of
the precision in the MOS calibration to absolute flux measurements, however,
a possible caveat could be larger uncertainty in the energy-dependent
vignetting effects, that are important at energies $>5$ keV.  Achievement of
the agreement in the MOS position with other calibration methods presented
in this paper is therefore of importance.

%

\subsection{Comparison of Methods}
The main method is important in that it defines rather well the energy
dependent effect of the vignetting but gave some concerns about the rather
large offset in telescope axis. The later methods give confidence that this
offset is real, and give similar magnitudes for the effect.

Table~\ref{offset} summarises the inferred axis shifts from these different
existing measurements.

\begin{table*}
\caption{Summary of the different axis displacement values 
for the EPIC cameras inferred with the various measurement techniques. 
Units are internal camera values of 0.05 arcsecond in ``Detector'' coordinates.}\label{offset} 
\begin{tabular}{l|rrrr} \hline\hline
Instrument &\multicolumn{4}{c}{ M e t h o d}\\
coordinate&G21.5-09/3C58&Source &Diffuse &Coma\\ 
          &             &Elongation&Background&\\ \hline
   pn DET X&$1300\pm300$&$1140\pm200$ & $1100\pm300$&$1243\pm30$\\
   pn DET Y&$450\pm300$&$340\pm200$&$ 400\pm300$&$402\pm30$\\
MOS-1 DET X&$200\pm300$&$-320\pm200 $&$   0\pm200$&$110\pm40$\\
MOS-1 DET Y&$-50\pm300$&$ 540\pm200$&$    0\pm200$&$-200\pm70$\\
MOS-2 DET X&$340\pm300$&$-340\pm200 $&$ 300\pm200$&$550\pm60$\\
MOS-2 DET Y&$-1300\pm300$&$-1700\pm200$&$ -1300\pm200$&$-1255\pm30$\\
\hline
\end{tabular}

\end{table*}

The discrepancy in axis value obtained by source elongation method for
the MOS cameras is partly explained by the differential shadowing of
RGA stack that modulates the PSF shape. this was modelled by ray trace
as described 
in section~\ref{sec:elong}. The RGA dispersion direction is parallel
with the MOS1 detector -Y axis, and the MOS2 detector +X axis. The
calculated shift due to differential shadowing amounts to
$\sim$440$\pm$20 units, which reduces the discrepancy in that axis to
$\sim$1$\sigma$. The discrepancy in the orthogonal direction is not
explicable by the ray trace, but is only $\sim$2$\sigma$. Additional
systematic uncertainty may be brought about by the non-circular shape
of the PSF (roughly pentangle and triangle shapes for MOS1 and MOS2
respectively) that is invariant with field angle and due to some
distortion encountered on mounting the mirrors to the spacecraft
interface plate.

\section{CONCLUSIONS}
\begin{figure}[ht]
  \begin{center}
   \begin{tabular}{c}
   \includegraphics[scale=0.5]{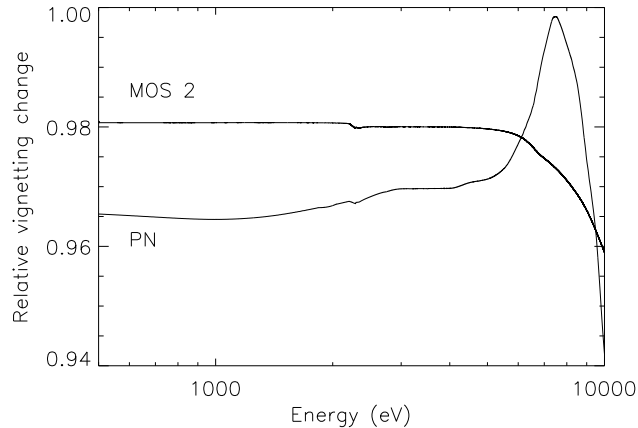}
\end{tabular}
  \end{center}
\caption{ Ratio of uncorrected and vignetted effective areas for an
  on-axis target. The flux differences for typical 0.5-2keV band will
  be $\sim$ a few \% for typical objects, and a small difference in recovered
  spectral slope will be caused by the change with energy  }\label{vigcal}
\end{figure}

The energy dependent vignetting calibration can be well matched to
pre-launch predictions, but only on an assumption that the telescope optical
axis is not perfectly aligned with the telescope boresight. This is not
unexpected following difficulties on-ground of maintaining and/or measuring
the telescope axis to better than 10's arcseconds.  We note that the assumed
telescope axis misalignment implies that ``on-axis'' targets at the common
boresight location are actually at a slightly different vignetting value per
telescope. We speculate that this partly accounts for some of the observed
flux discrepancies between the MOS and pn cameras (Figure~\ref{vigcal}). After reviewing
these data sets, it was decided that the XMM-Newton calibration
database would be updated in 2004 to account for new reference axes to
be centred at PN (DETX,DETY = 1240,400 ), MOS1 (DETX,DETY = 100,-200 )
and MOS2  (DETX,DETY = 500,-1250 ).

\section{Acknowledgments}
All the EPIC instrument calibration team who have contributed to
understanding the instrument are warmly thanked for their efforts. We thank
the XMM-Newton Science Operations Centre team for their help in scheduling
the calibration observations that needed special arrangements. AF
acknowledges receiving the Max-Plank-Gesellschaft Fellowship. {\em
  XMM-Newton} is an ESA science mission with instruments and
contributions directly funded by ESA Member States and the USA (NASA).

\end{article}
\end{document}